\documentclass[10pt, aps,prc,twocolumn,superscriptaddress,preprintnumbers,
amsmath, 
floatfix,
longbibliography,
nofootinbib
]{revtex4-1}
\usepackage[T1]{fontenc}
\usepackage[utf8x]{inputenc} 
\usepackage{adjustbox}          
\usepackage[caption=false]{subfig}
\usepackage{url}
\usepackage{color}
\usepackage{float}
\usepackage[pdftex,colorlinks=true, linkcolor = blue, citecolor=blue,urlcolor=blue, bookmarksnumbered=true, bookmarksopen=true]{hyperref}
\usepackage{longtable}
\usepackage{amsfonts}
\usepackage{wrapfig,bm}
\usepackage[normalem]{ulem}
\newcommand{\beq}{\begin{equation}}
\newcommand{\eeq}{\end{equation}}
\newcommand{\bea}{\begin{eqnarray}}
\newcommand{\eea}{\end{eqnarray}}

\begin{document}

\title{The angular correlation between the fission fragment intrinsic spins} 
  
\author{Aurel Bulgac }%
\email{bulgac@uw.edu}%
\affiliation{Department of Physics,%
  University of Washington, Seattle, Washington 98195--1560, USA}
   
\date{\today}

\begin{abstract}

The generation of fission fragments (FF) spins and of their relative
orbital angular momentum has been debated for more than six decades
and no consensus has been yet achieved so far. The interpretation of
recent experimental results of Wilson {\it et al.} [Nature, {\bf 590}, 566 (2021)] have been
challenged by several recent theoretical studies, which are not in
agreement with one another. According to  Wilson {\it et al.}'s 
interpretation [Nature, {\bf 590}, 566 (2021)], the FFs spins emerge very long after
scission occurred. Randrup and Vogt [Phys. Rev. Lett. {\bf 127}, 062502 (2021)], while agreeing that the 
FF spins are uncorrelated, conclude on the basis of a
phenomenological model that these spins are
uncorrelated already before scission. Bulgac {\it et al.} [Phys. Rev. Lett. {\bf 128}, 022501 (2022)] in a
fully microscopic study demonstrate
that the primordial FF spins final values are defined
before the emission of prompt neutrons and statistical gammas and are
strongly correlated with a relative angle between spins close to
$2\pi/3$, a result in full agreement with the present independent analysis.
The prompt neutrons and statistical gammas carry a
significant amount of angular momentum according to the study of
Stetcu {\it et al.} [Phys. Rev. Lett. {\bf 127}, 222502 (2021)], which can lead to a decorrelation of the
FF bandhead spins of the yrast lines measured by
 Wilson {\it et al.} [Nature, {\bf 590}, 566 (2021)], and which also provides arguments why these measured spins are so
different from the primordial FF spins evaluated by
Bulgac {\it et al.} [Phys. Rev. Lett. {\bf 128}, 022501 (2022)]. Here, I show that the unexpected character of
the angular correlation between the primordial FF
intrinsic spins, recently evaluated by Bulgac {\it et al.} [Phys. Rev. Lett. {\bf 128}, 022501 (2022)], 
which favors FF intrinsic spins pointing predominantly in opposite
directions, can be understood by using simple general phase space
arguments.  The observation by Wilson {\it et al.} [Nature, {\bf 590}, 566 (2021)] that the  FF spins 
are uncorrelated follows from both the results of the microscopic 
calculation of Bulgac {\it et al.}  [Phys. Rev. Lett. {\bf 128}, 022501 (2022)] and the present 
analysis of the full correlated probability distribution of the FF spins 
together with the relative orbital angular momentum of the primordial FFs. 
These arguments may apply also to heavy-ion 
collisions and since there is no use of specifics of the particle
interactions, the present results might apply to atomic and molecular
systems as well.

\end{abstract}

\preprint{NT@UW-21-11}

\maketitle

\section{Introduction}

The origin and the dynamics of the fission fragments (FFs) intrinsic spins
and of their relative orbital angular momentum between the FFs are 
topics with a very long history and many unsettled and conflicting
interpretations and claims, see a representative, but definitely not
complete, list of
references~\cite{Strutinsky:1960,Ericson:1960,Huizenga:1960,Vandenbosch:1960,Nix:1965,Rasmussen:1969,
Wilhelmy:1972,Vandenbosch:1973,Moretto:1980,Dossing:1985,Dossing:1985a,Moretto:1989,Wagemans:1991,Stetcu:2013,
Becker:2013,Randrup:2014,Vogt:2020,Randrup:2021,Wilson:2021,Bulgac:2021,Marevic:2021,Randrup:2021,Bulgac:2021b,Stetcu:2021}.
The results of the complex microscopic simulations performed in
Ref.~\cite{Bulgac:2021b} proved to be an
unexpected surprise, at odds both with the interpretation by
\textcite{Wilson:2021} of their experimental data, and also at odds
with a recent study performed by \textcite{Randrup:2021,Vogt:2020} within the
framework of the phenomenological model FREYA.  Moreover, while 
\textcite{Randrup:2021} agree with \textcite{Wilson:2021}
that the FF intrinsic spins are at most 
weakly correlated, in their analysis these FF spins emerge almost
uncorrelated \emph{before scission}, while the assumed nucleon-exchange mechanism  
between the preformed FFs is at work~\cite{Randrup:1979,Randrup:1982}. Thus, the analysis of
\textcite{Randrup:2021} invalidates the interpretation of
\textcite{Wilson:2021} of their own experimental results, who state
that the FF spins emerge uncorrelated \emph{after scission}, a claim
which cannot be reconciled with any conceivable theoretical
dynamic model. In a subsequent study within the Hauser-Feshbach
framework~\cite{Hauser:1952}, \textcite{Stetcu:2021}, demonstrate that
after scission, the highly excited primordial FFs emit prompt neutrons
and statistical gammas, which carry a significant amount of angular
momentum, $\approx 3.5-5\, \hbar$. These results can explain the observed
values of the yrast bandheads measured by \textcite{Wilson:2021} and
also why these FF spins might appear uncorrelated at that point in the 
fission dynamics.

Here I will present a transparent analysis of the unexpected 
theoretical prediction presented recently in
Ref.~\cite{Bulgac:2021b}, namely that the directions of the primary FF
intrinsic spins are strongly correlated, in a manner not inferred from
previous studies, either experimental, phenomenological, or
microscopic.  One can suspect that the complexity of the
implementation of the Time-Dependent Density Functional Theory
(TDDFT)~\cite{Shi:2020,Bulgac:2016,Bulgac:2019b,Bulgac:2020} could
contain some unidentified errors and subsequenlty lead to an erroneous
conclusion in Ref.~\cite{Bulgac:2021b}.  The uncertainties of the 
nuclear energy density functionals (NEDF)~\cite{Salvioni:2020} or
complexity of the TDDFT numerical implementation could hide some
erroneous inputs. One can also speculate that long memory
effects~\cite{Gross:2006,Gross:2012} are relevant when the collective
velocities of the fission dynamics are even slower than in the
adiabatic approximation~\cite{Bulgac:2019b,Bulgac:2020}.  On the other
hand phenomenological models typically rely on a large number of
parameters and nuclear properties, many of them not known with
sufficient precision, if at all. As the recent microscopic studies have
shown~\cite{Bulgac:2021,Bulgac:2021b,Marevic:2021}, phenomenological
studies~\cite{Litaize:2015,Verbeke:2018,Talou:2021,Schmidt:2016,Schmidt:2018,Vogt:2020}
assume incorrectly that the moment of inertia of the heavy (H) FF is
larger than that the light (L) FF, an aspect corrected
in Ref.~\cite{Randrup:2021}, where an {\it ad hoc} phenomenological parametrization 
of the FF moments of inertia was introduced. Microscopic
studies~\cite{Bulgac:2019b,Bulgac:2020} also clearly demonstrate that 
the FF temperatures are different and typically the heavy FF has less
excitation energy than the light FF, an aspect neglected in FREYE 
model~\cite{Vogt:2020,Randrup:2021}.  The interpretation of the
experimental results of \textcite{Wilson:2021} and the recent 
conflicting theoretical and phenomenological
conclusions~\cite{Bulgac:2021,Bulgac:2021b,Marevic:2021,Vogt:2020,Randrup:2021,Stetcu:2021},
and also the range of conflicting initial assumptions in these
approaches, might fail to convince the wider audience of their
reliability and can greatly benefit from an independent analysis.

\section{Theoretical approach}

The theoretical framework adopted here has similarities with the
Fermi's golden rule derivation, where the transition probability per
unit time from an initial state to all allowed by conservation laws
final states, is the product between the square of an average matrix
element, often taken as a phenomenological constant, and the density
of the final states, which can be often easily evaluated.  A
particular pedagogical example of the use of the Golden rule is that
of rate of the neutron $\beta$-decay, which shows that the shape of
the electron spectrum is fully determined by the density of final
states and the energy, momentum, and angular momentum conservation
laws.  

One can define the angle between the FF intrinsic spins,
\begin{align}
\phi^{LH} = \left \langle 
\arccos  \frac{{ \bm{S}}^{L}\cdot {\bm{S}}^{H} } { {S^{L}}{S}^{H} } 
 \right \rangle, 
 \label{eq:acos}
\end{align} 
where the brackets $\langle\, \rangle$ stand for the quantum mechanical expectation 
value of this complex operator.  I will show that
in the case of FF intrinsic spins, the shape of the distribution
$p(\phi^{LH})$ is controlled by the structure of the available
phase space of the final FF intrinsic spins and their relative orbital
angular momentum, and that this distribution is only weakly dependent
on the fission mechanism.  

The only input needed in the present analysis will be the quantum
theory of angular momentum and only some very mild and quite general
and flexible assumptions about the individual angular momentum
distributions. 
Since all relative FF intrinsic spin degrees of freedom, bending and
wriggling, twisting and tilting, are treated explicitly, the present
analysis is more general than the microscopic
treatment presented in Ref.~\cite{Bulgac:2021b} and 
FREYA~\cite{Randrup:2021,Vogt:2020}, where only the bending and
wriggling modes were explicitly considered.

I will limit at first the analysis to the very clean case of the
spontaneous fission of an even-even nucleus, such as $^{252}$Cf(sf),
which has an initial spin and parity $S_0^\pi=0^+$. In this case the
two FF intrinsic spins ${\bm S}^{L,H}$ and their relative orbital
angular momentum ${\bm \Lambda}$ satisfy the relation
 \begin{align} 
\langle  {\bm S_0} \rangle = \langle {\bm S}^{L} + {\bm S}^{H} +{\bm 
\Lambda}\rangle  ={\bm 0}, \label{eq:S0} 
\end{align} 
and where by definition $\Lambda$ is an integer. In the
classical limit all three components of each of these angular momenta
have well defined values. In this limit these three vectors lie
in a plane and ${\bm \Lambda}$ is perpendicular to the
fission direction.  The FF intrinsic spins ${\bm
S}^{L,H}$ are not restricted to lie in a plane perpendicular to
the fission direction in the present analysis, unless the FFs emerge 
only with $K^{L,H}=0$ projections of their respective intrinsic spins.  At the quantum
mechanical level only the magnitude and one cartesian component of
each angular momentum can have simultaneously well-defined
values. Quantum fluctuations therefore lead to ``fluctuations'' of the
orientation of the plane formed by these three vectors.  Before
scission the identity of the FFs is not uniquely defined, as matter, 
momentum, angular momentum, and energy is flowing between them. The FF
intrinsic spins and ${\bm \Lambda}$ are well-defined
only at a sufficiently large separation
between FFs, albeit for the role of the long-ranged Coulomb
interaction~\cite{Bulgac:2020a,Bertsch:2019a} and the 
presentations of Bertsch, Scamps, Bulgac at the recent {\it Workshop on Fission Fragment Angular Momenta}~\cite{workshop:2022}.

Since the average values of these
three angular momenta are ${\cal O}(10
\,\hbar)$~\cite{Bulgac:2021,Bulgac:2021b,Marevic:2021,Randrup:2021,Stetcu:2021,Vogt:2020}
in the following discussion I will use often use a classical
interpretation of the angular momenta. Eq.~\eqref{eq:S0} implies that out of the 9 
Cartesian components of these angular momenta only 6 are independent and 
I choose the triangle formed by them to lie in the $Oxy$-plane and the vector 
${\bm \Lambda}$ to be along the $Ox$-axis. In that case Eq.~\eqref{eq:S0} can be re-written explicitly as 
\begin{subequations}
\begin{align}
& S^{L}_x+S^{H}_x = -\Lambda, \label{eq:twist} \\
& S^{L}_y= - S^{H}_y ,\label{eq:wrigg}\\
&S^{L}_z= S^{H}_z = 0, 
\end{align}
\end{subequations}
and thus only 3 cartesian components are then left independent.
The fission direction then lies in the $Oyz$-plane.

\begin{figure}
\includegraphics[width=1.09\columnwidth]{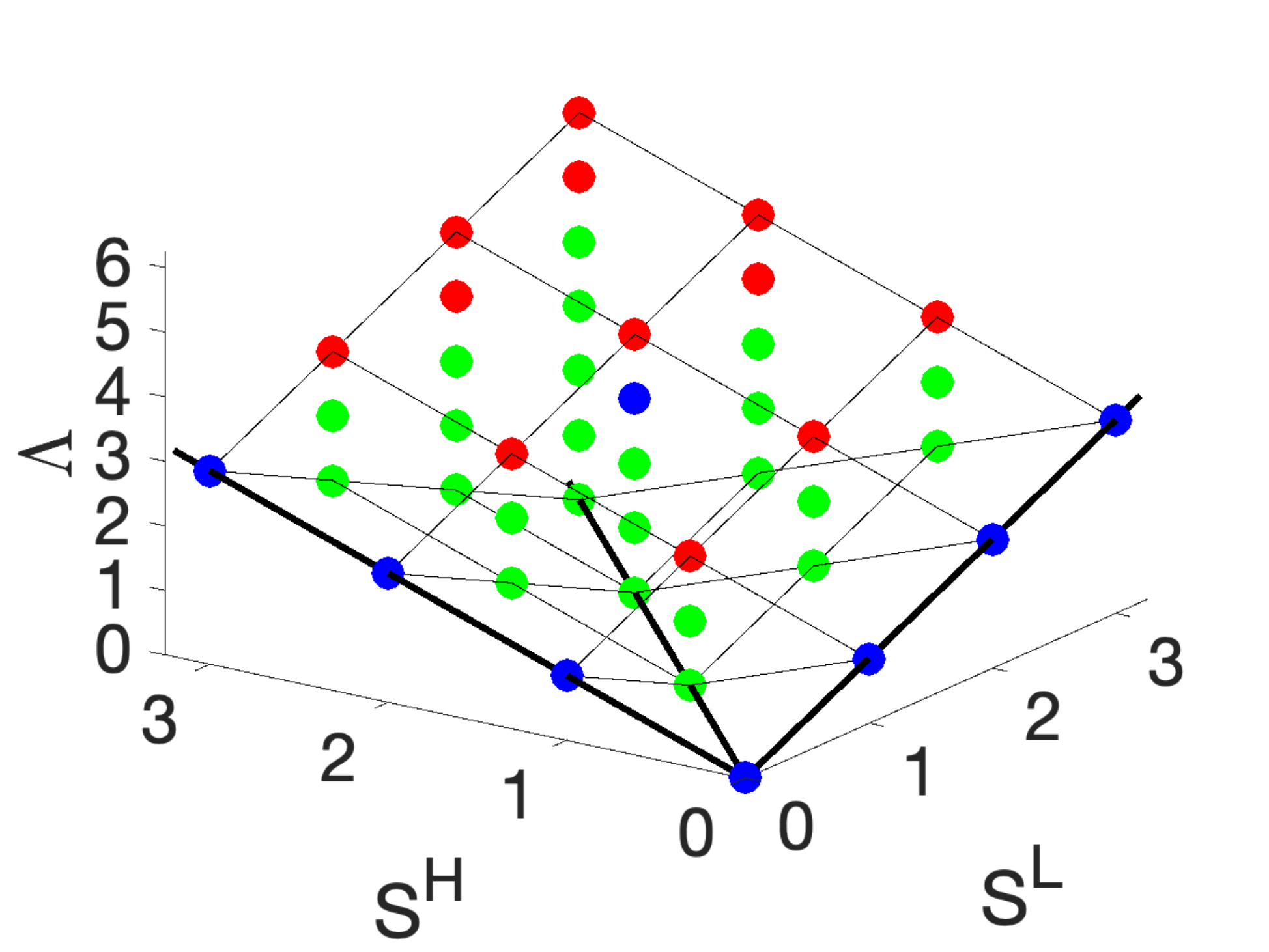}
\caption{ \label{fig:P3}  
In this figure the bullets fill out a
three face pyramid with the apex at (0, 0, 0).T
he green bullets show the triplets $(S^{L},S^{H},\Lambda)$
for which $\cos \phi^{LH} <0$, the blue bullets for $\cos
\phi^{LH} =0$, and the red bullets for $\cos \phi^{LH} >0$,
when $S_\text{max}=3$.  The green bullets correspond to 
$\phi^{LH}> \pi/2$ and the red bullets correspond to $\phi^{LH}<\pi/2$.
The ratio of red to green
bullets for any $S_\text{max}$ value is always close to 0.5, which
means that the number of configuration in which the FF intrinsic spins
point in opposite direction is dominant.  The black lines are the intersection 
of the pairs of planes $(S^{L}=S^{H}, \Lambda = 0)$, $(S^{L}=\Lambda,
S^{H}=0)$, $(S^{H}=\Lambda, S^{L}=0)$ respectively
and they cross at (0, 0, 0).  The points on each face are joined by thin lines. }
\end{figure}

I will choose the magnitudes of these angular momenta 
$\Lambda$ and $S^F =\sqrt{(S^F_x)^2+(S^F_y)^2}$ as variables in the following analysis. 
The magnitudes of these angular momenta satisfy the triangle restriction
\begin{align}
|{S}^{L}-{S}^{H}| \le {\Lambda} \le { S}^{L}+{ S}^{H}. \label{eq:3Js}
\end{align}
I will assume that there is an upper momentum cutoff $S_\text{max}$
for $S^{L,H}$, which can be taken to infinity if ever needed.
FFs can be odd or odd-odd nuclei, and in the case of odd FFs the
intrinsic spin is a half-integer. I ignore this possibility here,
which however is trivial to include and does not affect the
conclusions.  It is easy to see that the number of points in the
3-dimensional space consistent with this
maximum angular momentum is $(2S_\text{max}+1)(S_\text{max}+1)^2$.  At
the same time, the triangle constraint \eqref{eq:3Js} allows only for
about 1/3 of the total number of configurations, more exactly for
\begin{align}
N_0=\frac{2}{3}(S_\text{max}+1)^3 +\frac{1}{3}(S_\text{max}+1)\approx \frac{2}{3}S_\text{max}^3. \label{eq:N}
\end{align}
In Fig.~\ref{fig:P3} I plot the distribution of the allowed values of 
the triplet $(S^{L},S^{H},\Lambda)$ for a small value of
$S_\text{max}=5$. The allowed angular momenta $S^\text{L,R}$ and
$\Lambda$ are in the interior of a three faced pyramid defined by the planes 
\begin{subequations} \label{eq:planes} 
\begin{align}
+\Lambda -S^{L}-S^{H}=0,\label{eq:up}\\
-\Lambda +S^{L}-S^{H}=0,\label{eq:l1}\\
-\Lambda -S^{L}+S^{H}=0. \label{eq:l2}
\end{align}
\end{subequations}
The number of triplets $(S^{L},S^{H},\Lambda)$ corresponding
to $\phi^{LH} >\pi/2$ (green bullets) is about twice as large
than the number of points corresponding to $\phi^{LH} <\pi/2$
(red bullets) for any value of $S_\text{max}$ and that is the main
reason why the angles $\phi^{LH}>\pi/2$ between the two FF spins
are favored.  In the limit $S_\text{max}\rightarrow\infty$  a fraction
0.65 of the pyramid volume corresponds to $\phi^{LH} >\pi/2$ and 
there only less than a quarter of the allowed values for these 
angular momenta correspond to angles $\phi^{LH}< \pi/2$.

There is another very simply qualitative argument on why the angle
between the two FF intrinsic spins is larger than $\pi/2$.  The three
spins ${\bm S}^\text{L, H}$ and $\Lambda$ can be arranged tail to tail
to form a ``Mercedes''-like star. The triangle
constraint~\eqref{eq:3Js} is fully symmetric with respect to any 
permutation of the three angular momenta and can be re-written in two
other equivalent forms
\begin{subequations}\label{eq:3Ja}
\begin{align}
|{S}^{L}- {\Lambda}| \le{S}^{H} \le { S}^{L}+ {\Lambda}, \\
 |{S}^{H}- {\Lambda}| \le{S}^{L} \le { S}^{H}+ {\Lambda}, 
\end{align}
\end{subequations}
as it is clear as well from Eqs.~\eqref{eq:planes}. There is no 
preferential role for $\Lambda$ and on average the angle between any
two angular momenta is naturally $2\pi/3$ for any random arrangement
of them, if all three angular momenta are generated with identical
distributions.  
  
The arguments presented above refer only to the structure of the FF
intrinsic spin final configurations, when all three spins add to zero
and make no reference to possible role of the dynamics and
in this sense this argument is similar to the one used in deriving the
Fermi's golden rule.

\begin{figure}
\includegraphics[width=1\columnwidth]{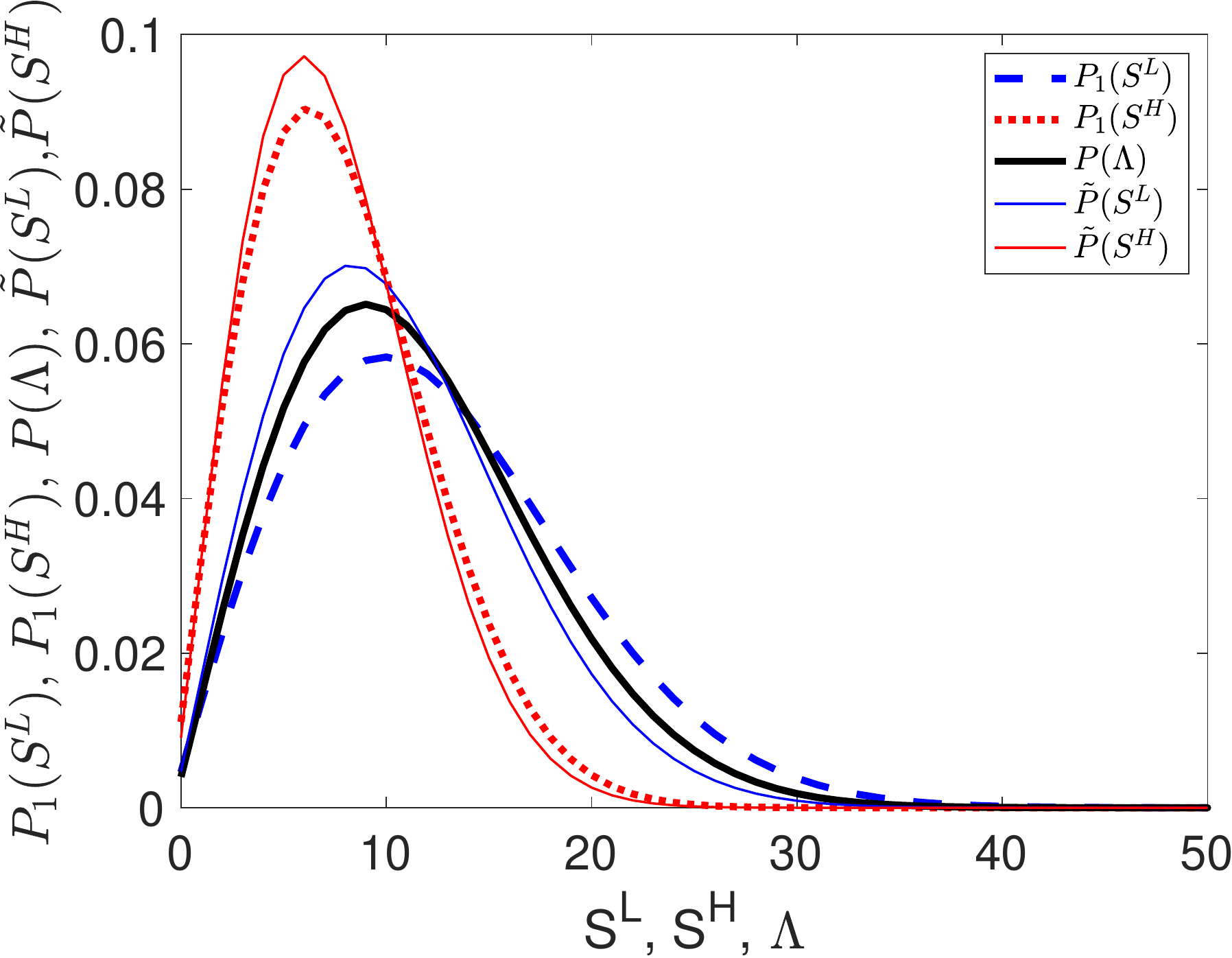} 
\caption{ \label{fig:TD1} 
The distributions $P_1(S^{L, H})$~\eqref{eq:S1} 
along with the distributions $\tilde{P}(S^{L, H})$ and $P(\Lambda)$,  see Eqs.~(\ref{eq:ss}, \ref{eq:LLL}) obtained form 
distribution $\overline{P}_3(S^{L},S^{H},\Lambda)$~\eqref{eq:TD} by summing over the rest of angular
momenta, are shown for $^{252}$Cf.  The FF intrinsic spin distributions $P_1(S^F) \approx \tilde{P}(S^F)$ and the 
relative orbital angular momentum distribution $P(\Lambda)$ obtained from 
Eq.~\eqref{eq:LLL} is very similar in shape to the statistical distribution  $P_1(\Lambda)$ from Eq.~\eqref{eq:L1}.
  }
\end{figure}

The next step is to generate distributions of these FF intrinsic spins
and orbital angular momentum consistent with Eq.~\eqref{eq:3Js}  
and their form depends on the nucleon interactions.
I will consider here at first two types of spin distributions, namely a
uniform distribution
\begin{align} 
P_1(S^{L,H}) = \frac{1}{S_\text{max}}, \quad P_1(\Lambda) = \frac{1}{2S_\text{max}}, \label{eq:uni}
\end{align}
which implies total ignorance about the angular momenta and a
statistical distribution, derived in the Fermi gas model by Bethe in
1936~\cite{Bethe:1936,Ericson:1960}
\begin{subequations} \label{eq:SLL} 
\begin{align} 
\!\!\!\!\!P_1(S^{L,H}) &\propto (2S^{L,H}+1) 
\exp \left [ -\frac{S^{L,H} (S^{L,H}+1)}{2\sigma^{L,H}}\right ], \label{eq:S1}\\
P_1(\Lambda)      &\propto (2\Lambda+1)      
\exp \left [ -\frac{\Lambda(\Lambda+1)}{2\sigma^\Lambda} \right ].  \label{eq:L1}
\end{align}
\end{subequations}  
The experience accumulated since 1936 is that a statistical
distribution based on the Fermi gas model is quite close to reality,
if one neglects quantum or shell effects.  Here, the parameters
$\sigma^{L,H}$ and $\sigma^\Lambda$ are chosen so as to reproduce
approximately the corresponding FF intrinsic spins distributions determined in
microscopic simulations~\cite{Bulgac:2021,Bulgac:2021b}.  Bethe's
approximation of a Fermi gas distribution seems quite reasonable,
apart from shell-corrections, as nucleons move predominantly in a mean
field, and level density and spin distributions are thus quite
reasonable. However, one cannot make the same argument concerning
$P_1(\Lambda)$. The ansatz 
\begin{align}
P(\Lambda) \propto \sum_{ S^{L,H} } P_1(S^{L}) 
P_1(S^{H}) {\bm \triangle} \label{eq:LL}
\end{align}
with $P_1(S^{L,H})$ 
from Eq.~\eqref{eq:S1} and where ${\bm \triangle}$ 
\begin{align}
\triangle = \Theta(\Lambda -S^L-S^H)\Theta(|S^L-S^H|-\Lambda)
\end{align}
with $\Theta(x)=1$ if $x\geq 0$ and $\Theta(x)=0$ if $x<0$ is enforcing the triangle rule, see 
Eqs.~(\ref{eq:3Js}), 
leads to a distribution $P(\Lambda)$ very similar to Eq.~\eqref{eq:L1}. 
See also the discussion 
below and later on see Eqs.~\eqref{eq:TD}, (\ref{eq:LLL}) and Fig.~\ref{fig:TD1}, for a 
different choice for $P(\Lambda)$ suggested by Thomas Dossing at 
the recent {\it Workshop on Fission Fragment Angular Momenta}~\cite{workshop:2022}. 

From such individual FF intrinsic spins and orbital angular momentum
distributions I generate the combined distributions
\begin{align}
 \overline{P}_3(S^{L},S^{H},\Lambda)= {\cal N}P_1(S^{L}) 
 P_1(S^{H}) P_1(\Lambda) {\bm  \triangle}, \label{eq:PPP}
 \end{align}
where 
\begin{align}
&\sum_{S^{L,H},\Lambda} \overline{P}_3(S^{L},S^{H},\Lambda)=1,
\end{align}
and where ${\cal N}$ is an appropriate normalization factor.  Alternatively, one can use also
$P_1(\Lambda)$ from Eq.~\eqref{eq:LL} with little changes in the
final results.  This combined distribution $\overline{P}_3(S^{L},S^{H},\Lambda)$ 
vanishes outside the region
defined by Eq.~\eqref{eq:3Js} in the 3-dimensional space
$(S^{L} ,S^{H},\Lambda)$ and it allows 
 for the presence of the twisting and tilting modes.
 
In the simulations each of the FF intrinsic spins and of the orbital
relative angular momentum populate large angular momentum
intervals. It is important to recognize that Eq.~\eqref{eq:3Js}
implies that the support of these distributions cannot be drastically
different, as otherwise Eqs.~(\ref{eq:3Js}, \ref{eq:3Ja}) cannot be
satisfied. If one of these distributions, e.g. $P_1(\Lambda)$, is very
wide, while the other two are much narrower, the triangle constraint
can be satisfied only for relatively small values of the angular
momenta $\Lambda$, and the longer tails of $P_1(\Lambda)$ will never
contribute to any physical situation, see also the discussion below.
Therefore, such drastically different angular momentum distribution
cannot emerge from any realistic calculations, where rotational
symmetry is satisfied.  
 
 In this work I determine the probability 
distribution $p(\phi^{LH})$, $\int_0^\pi d\phi^{LH}
p(\phi^{LH})=1$, where $\phi^{LH}$ is the angle between
$S^{L}$ and $S^{H}$ by constructing a histogram of the
expectation of the cosine between them
\begin{align}
\!\!\!\cos \phi^{LH} = \frac{ \Lambda(\Lambda+1) - 
S^{L}(S^{L}+1) -S^{H}(S^{H}+1)}{2(S^{L}+1/2)(S^{H}+1/2)}. \label{eq:cos}
\end{align}
The magnitudes of the 3 angular momenta are expressed in units of
$\hbar$ and I will use the units $\hbar =1$ henceforth.  In the
denominator of Eq.~\eqref{eq:cos} I introduced the well known Langer 
correction~\cite{Langer:1937} to the FF intrinsic spins, which is not
needed in the numerator. Eq.~\eqref{eq:cos} can be readily derived
from Eq.~\eqref{eq:S0} by replacing the scalar product with
\begin{align}
{\bm S}^{L}\cdot{\bm S}^{H}\rightarrow (S^{L}+1/2)(S^{H}+1/2) \cos \phi^{LH}.
\end{align}
The angle $\phi^{LH}$ depends only on the shape
of the triangle, or the shape of the ``Mercedes-star'' the three
angular momenta form, but not on its spatial size  
nor on its particular orientation in space. The only relevant dimensional 
scale, $\hbar$, vanishes in the
classical limit and as a result all quantities ${\cal O}(1)$ should be
neglected in this limit, e.g. 1/2 and 1 in Eq.~\eqref{eq:cos}.  

\begin{figure}[h]
\includegraphics[width=1\columnwidth]{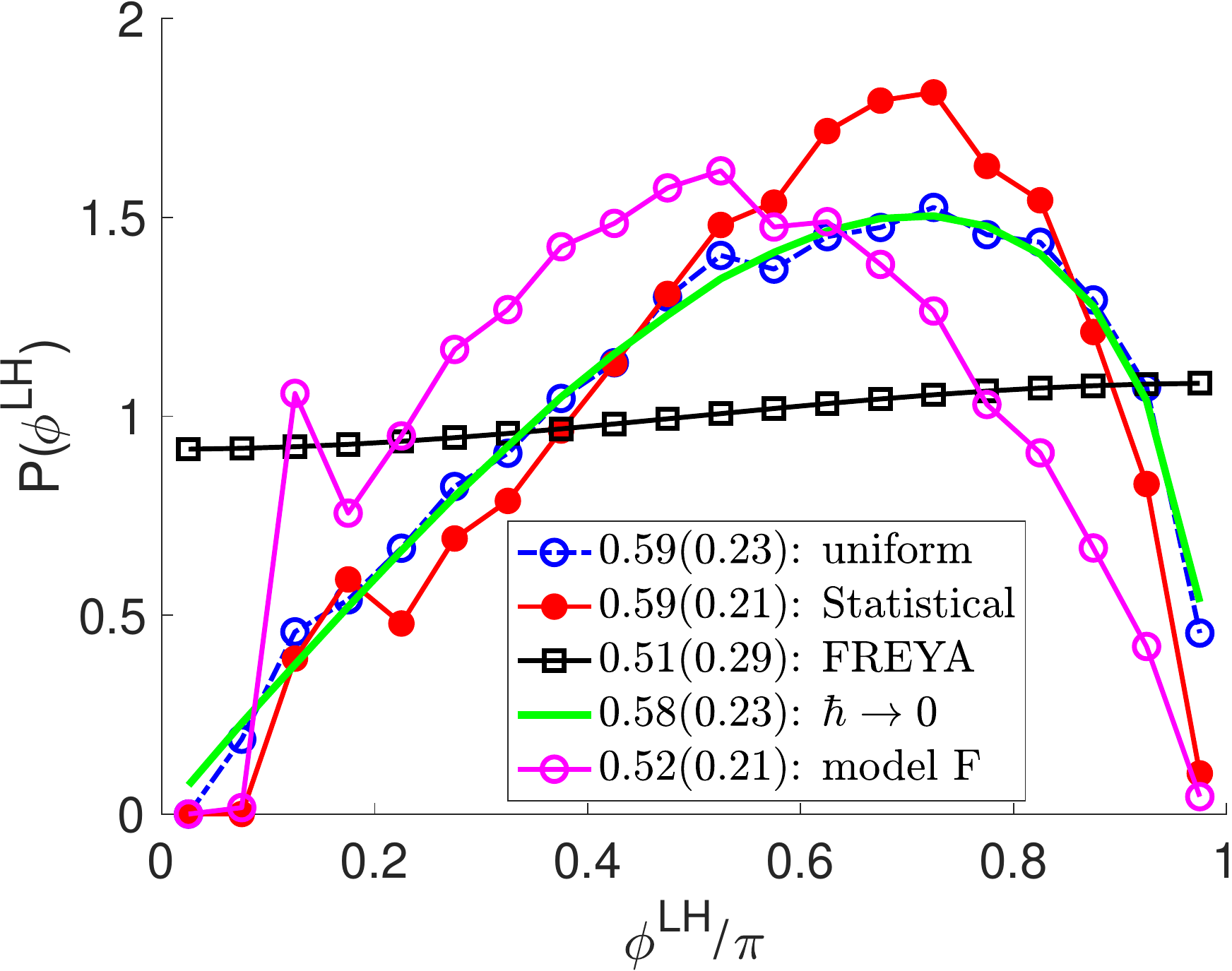} 
\caption{ \label{fig:P_phi} The uniform angular momentum distribution 
\eqref{eq:uni} was obtained with an $S_\text{max}= 30$, see
Eq.~\eqref{eq:uni}, and represented here as a histogram with 20 bins.
The statistical distribution was obtained with $\sigma^{L} =
108.2,\, \sigma^{H}= 44.8$ and $\sigma^\Lambda =161.3$, which
closely reproduce the corresponding distributions obtained in
Refs.~\cite{Bulgac:2021,Bulgac:2021b} for $^{240}$Pu. The FREYA distribution was
obtained with typical parameters 
from~\cite{Vogt:2020,Randrup:2021}. The $\hbar\rightarrow 0$ classical
limit was obtained by taking $S_\text{max} \rightarrow\infty$ for the
uniform distribution.  The results obtained with a modified FREYA
prescription $P_\text{model F}$ introduced here are shown
with magenta circles.  The average and variance (in parentheses) of
$\phi^{LH}/\pi$ for each distribution are displayed next to each
label.  The lines connecting the markers are only for guiding the eye.
For visual purposes the distribution displayed here differs by a
factor from Eq.~\eqref{eq:pp}, $P(\phi^{LH})=\pi p(\phi^{LH}).$
The FREYA results from Ref.~\cite{Randrup:2021} 
and reproduced here are markedly different from the distributions 
derived here and from the microscopically evaluated distribution
$p(\phi^{LH})$ in Ref.~\cite{Bulgac:2021b}. 
In the space $(S^L,S^H, \Lambda)$ the allowed 
configurations with $\phi^{LH}=\pi$  correspond to 
the lateral lower faces of the pyramid in Fig.~\ref{fig:P3}, see Eqs.~(\ref{eq:l1}, \ref{eq:l2}),
when $\Lambda = |S^L-S^H|$.
The allowed configurations with $\phi^{LH}=0$ correspond to upper face of the pyramid  
$\Lambda = S^L+S^H$  in Fig.~\ref{fig:P3}, see Eq.~\eqref{eq:up}.
All faces  have dimension 2, as opposed to the dimension 3 
of the space of the rest of  allowed configurations $(S^L,S^H,\Lambda)$,
which makes obvious the presence of the strong suppression at $\phi^{LH}= 0,\, \pi$.
This clear discrepancy between 
the predictions made in Refs.~\cite{Vogt:2020,Randrup:2021} and reproduced 
here and the predictions made in the present work and in  
Ref.~\cite{Bulgac:2021b} remains to be resolved in future experiments, 
see also Sobotka's talk at the 
 {\it Workshop of Fission Fragment Angular Momenta}~\cite{workshop:2022}.   }
\end{figure}

I generate all $N_0$, see
Eq.~\eqref{eq:N}, triplet configurations
$(S^{L},S^{H},\Lambda)$ allowed by Eq.~\eqref{eq:3Js} and 
each such configuration is weighted with probability
$\overline{P}_3(S^{L},S^{H},\Lambda)$~\eqref{eq:PPP}. Then the
probability distribution $p(\phi^{LH})$ is constructed as follows
\begin{align} 
&p(\phi^{LH}) = \sum_{S^{L},S^{H},\Lambda}  \overline{P}_3(S^{L},S^{H},\Lambda) 
\delta_{\psi(S^{L},S^{H},\Lambda),\phi^{LH}}, \label{eq:pp}\\
&\sum_{\phi^{LH}=0}^{\phi^{LH}=\pi}  p(\phi^{LH}) =1,\\
&\psi(S^{L},S^{H},\Lambda) \\
& = \arccos\left [ 
\frac{ \Lambda(\Lambda+1) - S^{L}(S^{L}+1) 
- S^{H}(S^{H}+1)}{2(S^{L}+1/2)(S^{H}+1/2)}\right ], \nonumber
\end{align}
where $\delta_{\psi,\phi}=1$ if $\phi=\psi$ and zero otherwise, is the
Kronecker symbol and the values
$\psi(S^{L},S^{H},\Lambda)$ are discrete.  Thus
$p(\phi^{LH})$ represents a discrete point with coordinates
$(\phi^{LH}, p(\phi^{LH}))$.  For each triplet
$(S^{L},S^{H},\Lambda)$ the corresponding angle
$\psi(S^{L},S^{H},\Lambda)$ is weighted with probability 
$\overline{P}_3(S^{L},S^{H},\Lambda)$ and its contribution is added to
the corresponding $\phi^{LH}$-bin in Fig.~\ref{fig:P_phi}.

In Fig.~\ref{fig:P_phi} I show five distributions for
$p(\phi^{LH})$ obtained by using the uniform distributions
Eq.~\eqref{eq:uni} (blue circles), the statistical distributions
Eqs.~\eqref{eq:SLL} (red bullets), the distribution predicted by the
FREYA model~\cite{Vogt:2020,Randrup:2021} (black squares), the
limiting classical distribution ($\hbar \rightarrow 0$) obtained by
taking the limit $S_\text{max}\rightarrow \infty$ (green line) for a
uniform distribution, and a model F distribution inspired by FREYA,
which is discussed below.

In the classical limit one can eliminate the angular momenta lengths
and introduce their relative normalized lengths
\begin{align}  
&s^{L}+s^{H}+\lambda =1, \label{eq:equi}\\
& \psi (s^{L},s^{H},\lambda)   
         = \arccos\frac{  \lambda^2 -(s^{L})^2-(s^{H})^2 }{2s^{L}s^{H}},
\end{align}
 and the   distribution $p(\phi^{LH})$ can be evaluated from

\begin{align}
 p(\phi^{LH}) &= 3\int_0^1 ds^{L}\int_0^{s^L}ds^{H}\int_{s^L-s^H}^{s^L+s^H} d\lambda \label{eq:class} \\
  &\times\delta [\phi^{LH} -\psi (s^{L},s^{H},\lambda)],  \nonumber
\end{align}
 when all shapes are equiprobable (uniform distributions).
 
The uncanny similarity to the distribution $p(\phi^{LH})$
reported in Ref.~\cite{Bulgac:2021b} is striking.  The distributions
in Fig.~\ref{fig:P_phi} were obtained without any input from  the dynamics, 
unlike the results of
Ref.~\cite{Bulgac:2021b}. By changing the form of the individual 
distributions $P_1(S^{L}) P_1(S^{H}) P_1(\Lambda)$ the final
aspect of $p(\phi^{LH})$ changes very little from the uniform to
the statistical distribution.  In the classical limit, which can be
achieved either for very large momenta or very high temperatures, the
distribution becomes very smooth (no quantum or ``shell-like
corrections''), as expected.  The triangle constraints~(\ref{eq:3Js},
\ref{eq:3Ja}) which enforce the conservation of the total angular
momentum of the entire nuclear systems leads to strongly correlated FF
intrinsic spins, as it is clear from Fig.~\ref{fig:P_phi}, with some
weak ``shell correction.'' As in the case of Bethe level density, the
role of the nucleon interactions is relatively small.  
The uniform angular momenta
distributions ~\eqref{eq:uni}, the statistical distributions
Eqs.~\eqref{eq:SLL}, and the classical limit \eqref{eq:class}, are
rather close to each other.  This implies that the primary FF
intrinsic spins form preferentially and angle larger than $\pi/2$
(similarly to any two prongs of a ``Mercedes-star"), thus 
favoring the bending modes over the wriggling modes.  There are no
qualitative changes when different distributions, because 
the number of configurations with $\phi^{LH}> \pi/2$ is always
about twice as big as the number of configurations with
$\phi^{LH}< \pi/2$, see Fig.~\ref{fig:P3}, which is the defining
feature of this distribution.  With increasing temperature, or
equivalently, in the classical limit, the size of the triangles
increases, but the shape remains the same, and the $p(\phi^{LH})$
changes relatively little, apart from the diminishing role of the
${\cal O}(\hbar)$ corrections, as expected on general grounds.  The
fact that one does not see a perfect agreement with the simple
geometrical arguments presented above, when the expected angle between
any two angular momenta is $2\pi/3$, is due to the fact that the
widths of the $P_1(S^{L,H})$ and $P_1(\Lambda)$ are different and
the full permutation symmetry between all these angular momenta is
slightly broken. In particular, in fission this symmetry is broken by the
mass and deformation differences between the heavy and light FFs.  

One can change the character of this FF intrinsic spins distribution
only by choosing a distribution of angular momenta
$\overline{P}_3(S^{L},S^{H},\Lambda)$, which drastically favors angles
$\phi^{LH}< \pi/2$, for example choosing a distribution
$P_1(\Lambda)$ which favors $S_\text{max} \le \Lambda \le
2S_\text{max}$. In such a situation the two FFs emerge at scission as
a system similar to a planet and its moon of comparable mass, rotating
around their common center-of-mass with a quite high frequency and
predominantly parallel spins. In such a case the probability
distribution $p(\phi^{LH})$ will favor angles $\phi^{LH} < 
\pi/2$ and the tilting and wriggling modes will be the dominant ones,
as opposed to the twisting and bending modes otherwise.  One can
imagine such a quasi-fission process in a heavy-ion collision with a
relatively large initial spin $S_0$ of the compound nucleus.  Since
the initial spin of the compound nucleus is now $S_0\neq 0$ one has to
discuss instead the probability distribution
$\overline{P}_3(S^{L},S^{H},|{\bm \Lambda}-{\bm S}_0|)$, which will
proceed along the same lines as above.

The distributions discussed here have a very distinct fingerprint,
they vanish in the classical limit at $\phi^{LH}=0$ and $\pi$, which happens in
Fig.~\ref{fig:P3} on the faces of the 
three face pyramid, where the allowed phase space is restricted to a 2D region. 
This feature is absent in FREYA
results~\cite{Randrup:2021,Vogt:2020}.

In FREYA the rotational energy, which controls the spin distributions, 
has the form~\cite{Vogt:2020,Randrup:2021}
\begin{align}
E_\text{rot}=  
                 \frac{ {\bm S}^{L}\cdot {\bm S}^{L}} {  2I^{L} } + 
                  \frac{ {\bm S}^{H}\cdot {\bm S}^{H} }{ 2I^{H} }
                  +\frac{{\bm \Lambda}\cdot {\bm \Lambda} }{2I^\text{R}},
                  \label{eq:rot}
\end{align} 
with the moments of inertia $I^\text{R}\gg I^{L,H}$.  This 
distribution can be used in a statistical description where $T$ is a
phenomenological temperature of the fissioning nucleus and $Z$ an
appropriate normalization factor.
\begin{align}
P(S^{L},S^{H},\Lambda)&=\frac{1}{Z} \exp \left [ -\frac{E_\text{rot}}{T} \right ], \nonumber \\ 
&=P_1(S^{L}) P_1(S^{H})P_1(\Lambda). \label{eq:F} 
\end{align}
This formula assumes that all these momenta can change only by
interacting with the rest of the degrees of freedom, which play the
role of a thermal bath, which equally implies that the two FFs are
still in contact with each other. Microscopic 
studies show~\cite{Bulgac:2019b,Bulgac:2020} that the FFs have however
different temperatures and it is unclear how a temperature for
$\Lambda$ can be defined then. 
While this ansatz for $E_\text{rot}$~\eqref{eq:rot} appears as a
natural starting point~\cite{Moretto:1980,Moretto:1989}, it does not
have the most general form allowed by rotational symmetry.  The
assumption that the FF shapes and their relative orientations do not
play any role in their dynamics appears to be very unlikely,
see some of the arguments presented by Bulgac, Bertsch and Scamps at the recent 
 {\it Workshop on Fission Fragment Angular Momenta}~\cite{workshop:2022} and Ref.~\cite{Bertsch:2019a}.  
 Assuming that one can limit the description of the rotation dynamics to these
three angular momenta alone, after eliminating all the relevant
angles, the most general form allowed by symmetry for the rotational 
energy is
\begin{align} 
E_\text{rot} = ({\bm S}^{L},{\bm S}^{H},{\bm \Lambda})^\text{T}  \otimes \tensor {\bm I}   
\otimes ({\bm S}^{L},{\bm S}^{H},{\bm \Lambda}), \label{eq:rott}
\end{align} 
with a non-diagonal $3\times 3$ inertia tensor $\tensor{\bm I}$ in
general, see also Ref.~\cite{Bulgac:2021b}.  A very simple analog is the effective
position dependent nucleon mass in mean field approaches, which in
principle can become a tensor as well~\cite{Bulgac:1995}, and also 
the numerous examples Brillouin zones with
strongly anisotropic electron kinetic energy dispersion.   

With no triangle constraints~(\ref{eq:3Js}, \ref{eq:3Ja}) and no
intrinsic spins correlation Eq.~\eqref{eq:F} implies $\langle {\bm
S}^{L}\cdot{\bm S}^{H}\rangle\equiv0$.  
Following the framework described above I introduce instead a
distribution inspired by FREYA
\begin{align}
P_\text{model F}(S^{L},S^{H},\Lambda)={\cal N}\exp\left [-\frac{E_\text{rot}}{T}\right ]\triangle
 \label{eq:Fmod}
\end{align} 
where ${\cal N}$ is an appropriate normalization factor and 
using Eq.~\eqref{eq:rot}.
This distribution is more general than the one used in the FREYA
phenomenological model, as it allows also for the tilting and twisting
degrees of freedom to be taken into account, and moreover at the
quantum level.  Even though this $P_\text{model
F}(S^{L},S^{H},\Lambda)$ depends on the initially very wide
distribution $P_1(\Lambda)$, see Eq.~\eqref{eq:F} with $E_\text{rot}$ 
from Eq.~\eqref{eq:rot}, the triangle constraint and the
subsequent renormalization lead to a physically acceptable 
distribution, with a physically acceptable average $\langle
\Lambda\rangle$ and variance/cumulant $\langle\langle
\Lambda^2\rangle\rangle$, due to similar arguments presented in
connection with Eq.~\eqref{eq:LL}.  In the limit of infinite
temperature the distribution $p(\phi^{LH})$ is expected to become
very smooth (no quantum or ``shell corrections'') similar to the
classical distribution shown with a green line in
Fig.~\ref{fig:P_phi}, for either the uniform~\eqref{eq:uni} or the
statistical distributions~\eqref{eq:SLL}, where I used
$S_\text{max}=200$.  
One can visually see a relatively small preponderance
of angles $\phi^{LH}> \pi/2$, as expected.  With increasing
temperature for this model F distribution the average and variance of
the angle $\phi^\text{HL}$ does not change significantly if
$I^\text{R}\gg I^{L,H}$.

In FREYA~\cite{Vogt:2020,Randrup:2021} the non-conservation of the
total angular momentum due to the presence 
of a thermal bath of the probability distribution~\eqref{eq:F} is resolved by
imposing the additional constraint $\Lambda = |{\bm S}^{L}+{\bm
S}^{H}|$.   All the angular momenta are treated classically ${\bm
S}^{L,H} \cdot {\bm S}^{L,H}=(S^{L,H})^2$ and the
tilting and twisting intrinsic spin degrees of freedom (the FF 
rotations around the fission axis) are assumed to be
suppressed~\cite{Vogt:2020,Dossing:1985}.  With increasing temperature
the angular momenta increase in size $\propto \sqrt{T}$ and since the
geometry of the triangle \eqref{eq:3Js} is the same at all
temperatures, the FREYA prediction~\cite{Vogt:2020,Randrup:2021} for
$p(\phi^{LH})$ is independent of temperature and FREYA framework and 
it is technically a high-temperature limit of the quantum distribution.
The primordial FF intrinsic spin correlations arise from the orbital
rotational energy term, see Eq.~\eqref{eq:rot}, only after the
replacement
\begin{align}
\frac{ {\bm \Lambda}\cdot{\bm \Lambda}}{2I^\text{R}} \rightarrow
\frac{{\bm  S}^{L} \cdot {\bm S}^{L}}{2I^\text{R}} 
+ \frac{{\bm  S}^{H} \cdot {\bm S}^{H}}{2I^\text{R}} +
\frac{{\bm  S}^{L} \cdot {\bm S}^{H}}{I^\text{R}} ,
\end{align}
and specifically from the last term in this relation  
with $I^\text{R}\approx 10 I^\text{L, H}$, which leads to very 
weak antiparallel FF intrinsic spins correlations, at the level of $\approx 10\%$, 
see Fig.~\ref{fig:P_phi} and Refs.~\cite{Vogt:2020,Randrup:2021}. 
An almost uniform $p(\phi^{LH})$ distribution as predicted by FREYA can be 
obtained in the unlikely case when the correlated distribution $P_3(S^L,S^H,\Lambda)$
diverges for $\Lambda =|S^L\pm S^H|$, see Appendix.

\begin{figure}
\includegraphics[width=1\columnwidth]{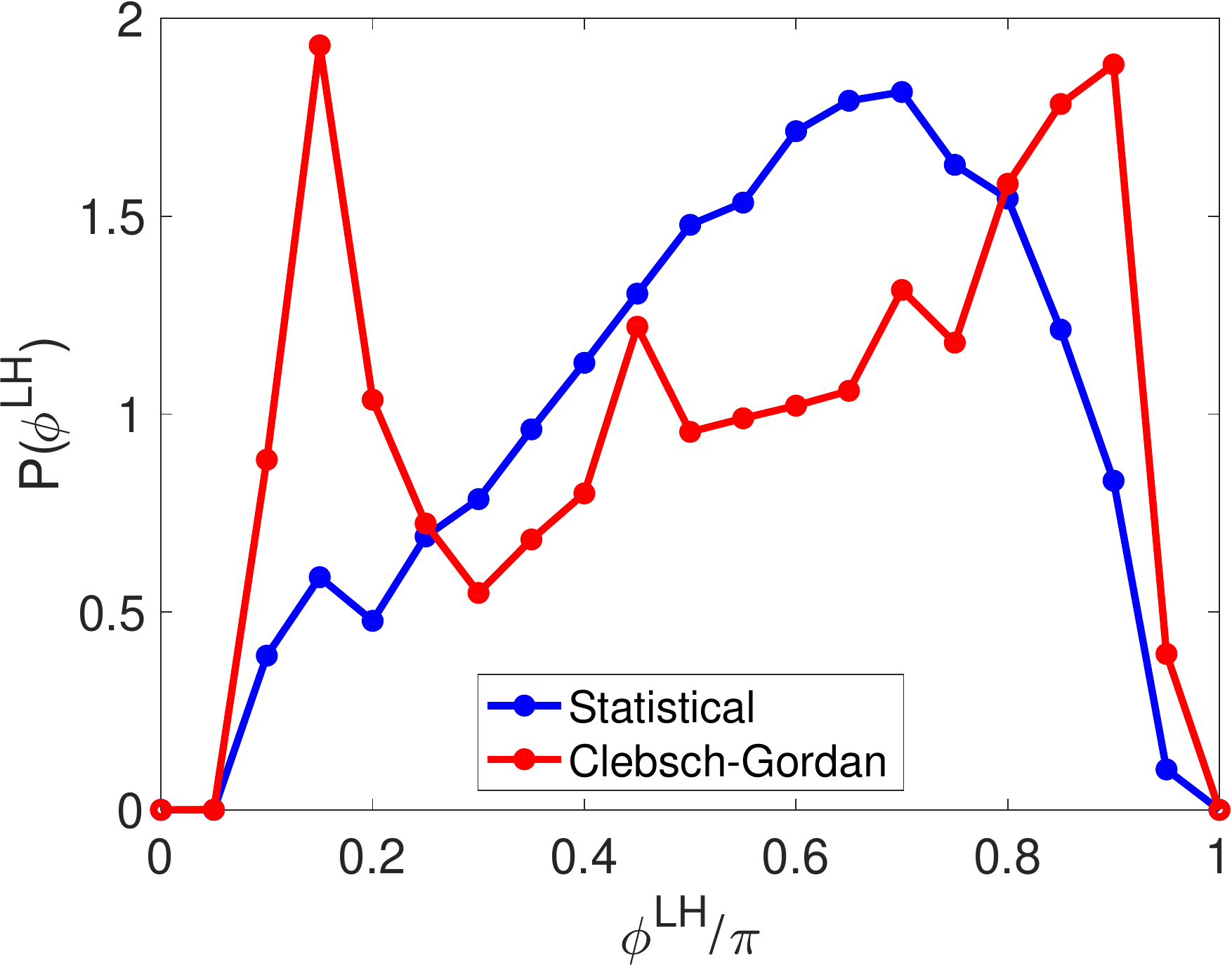} 
\caption{ \label{fig:TD2} 
The distribution  $P(\phi^{LH})$ (normalized as in Fig.~\ref{fig:P_phi}) obtained using the distributions
$\overline{P}_3(S^{L},S^{H},\Lambda)$ using Eq.~\eqref{eq:PPP} (Statistical) and 
$\tilde{P}_3(S^{L},S^{H},\Lambda)$ using Eq.~\eqref{eq:TD} (Clebsch-Gordan).
Both these distributions show a preference for angles $\phi^{LH}> \pi/2$ and both 
of them show a clear suppression of the FF intrinsics spins either parallel or 
anti-parallel to each other.  The statistical distribution here is slightly different from the one shown in Fig.~\ref{fig:P_phi}, 
since in that case I used $P_1(\Lambda)$ from Eq.~\eqref{eq:L1} while here I used $P(\Lambda)$ from Eq.~\eqref{eq:LLL}. 
 }
\end{figure}

During the recent {\it Workshop of Fission Fragment Angular Momenta}~\cite{workshop:2022}
the character of the distribution $p(\phi^{LH})$ was extensively discussed, 
see in particular the slides and the video recordings  
of the talks given by  Randrup, Vogt, Bulgac (2), Dossing and Sobotka. 
Thomas Dossing suggested an alternative parametrization of the 
probability distribution $\tilde{P}_3(S^{L},S^{H},\Lambda)$
\begin{align}
\tilde{P}_3(S^{L},S^{H},\Lambda)& = {\cal N}P_1(S^{L})P_1(S^{H}) \nonumber \\
&\times \left [C_{S^{L},0,S^{H},0}^{\Lambda,0}\right ]^2 \exp\left [ -\frac{\Lambda(\Lambda+1)}{2I_{rel}T} \right ] , \label{eq:TD}
 \end{align}
 where ${\cal N}$ is an appropriate normalization constant and $C_{S^{L},0,S^{H},0}^{\Lambda,0}$ are Clebsch-Gordan coefficients. 
 This distribution $\tilde{P}_3(S^{L},S^{H},\Lambda)$
 assumes that the primordial FF intrinsic spins, before the emission 
 of prompt neutrons and statistical $\gamma$-rays  are strictly perpendicular to the 
 fission axis ($K^{L,H}=0$), thus allowing only for the bending and wriggling modes of FF intrinsic spins, similarly to what it is 
 assumed in  FREYA~\cite{Vogt:2020,Randrup:2021}. 
 In the analysis of this probability distribution I have used the exact values of the 
 Clebsch-Gordan coefficients~\cite{Varshalovich:1989}. 
 It also assumes that the relative orbital angular momentum 
 is thermalized to the same temperature as the FF intrinsic spins. Randrup in his presentation at the 
 {\it Workshop of Fission Fragment Angular Momenta}~\cite{workshop:2022}
 stressed that FREYA can be also used with different temperatures for different rotational modes
 modes (wriggling, bending, and maybe twisting, which is typically neglected). 
 Here $I_{rel}$ was chosen as the relative moment of inertia 
 of two touching rigid sphere with the average mass close to the 
 average fission yields for induced fission of $^{240}$Pu and $T=1$ MeV. A
 fairly similar distribution  is obtained for either $T= 0.75$ or 0.5 MeV temperatures. 
 One can include the FF deformations if needed, see Ref.~\cite{Randrup:2021}, but I neglect this aspect here, as this is less relevant for $I_{rel}$.
 This distribution also favors slightly large relative   
 orbital angular momenta (with an average of $\langle\phi^{LH}\rangle = 0.60$
 and standard deviation 0.26), thus in agreement with Ref.~\cite{Bulgac:2021b}  
 and the present analysis, and in disagreement with the results of Refs.~\cite{Vogt:2020} and \cite{Randrup:2021}.
 The individual angular momenta distributions
 $ \tilde{P}_1(S^{L}),\tilde{P}_1(S^{H})$ and $\tilde{P}_1(\Lambda)$ extracted by summing over the rest 
 of the angular momenta, and which are shown in Fig.~\ref{fig:TD1} along 
 with the distributions from Eq.~\eqref{eq:S1} for $\tilde{P}_1(S^{L, H})$  
 \begin{align}
\tilde{P}_1(S^{F} )&= \sum_{\Lambda, f\neq  F} \tilde{P}_3(S^{F},S^{f},\Lambda),\label{eq:ss}\\
{P}(\Lambda) &= \sum_{S^L,S^H}\tilde{P}_3(S^{L},S^{H},\Lambda), \label{eq:LLL}
\end{align} 
are qualitatively similar, though not identical.
The probability distributions $p(\phi^{LH})$ extracted using the triple distributions 
 $\overline{P}_3(S^L,S^H,\Lambda)$ from Eq.~\eqref{eq:PPP} and $\tilde{P}_3(S^L,S^H,\Lambda)$ form Eq.~\eqref{eq:TD}, 
 as expected favor angles $\phi^{LH}>\pi/2$, see Fig:~\ref{fig:TD2}.

During the workshop~\cite{workshop:2022} I realized that from the microscopic
distribution $P_{\text{true}\,3}(S^{L},S^{H},\Lambda)$ from Ref. ~\cite{Bulgac:2021b} 
as well as from the distributions 
$\overline{P}_3(S^{L},S^{H},\Lambda)$ using Eq.~\eqref{eq:PPP} or
$\tilde{P}_3(S^{L},S^{H},\Lambda)$ using Eq.~\eqref{eq:TD}
one can extract the double FF 
intrinsic spin distributions. E.g. in the case of $\tilde{P}_3(S^{L},S^{H},\Lambda)$
\begin{align} 
\!\!\! \tilde{P}_2(S^{L},S^{H})&= \sum_\Lambda \tilde{P}_3(S^{L},S^{H},\Lambda)
\approx  \tilde{P}_1(S^{L})\tilde{P}_1(S^{H}), \label{eq:P2}
\end{align}
and similarly for the corresponding distributions $\overline{P}_2(S^{L},S^{H})$ obtained form $\overline{P}_3(S^{L},S^{H},\Lambda)$
and $P_2(S^{L},S^{H})$ obtained from $P_{\text{true}\,3}(S^{L},S^{H},\Lambda)$.
This proves  that the primordial FF intrinsic spins are essentially 
statistically independent
before the emission of neutrons and statistical $\gamma$-rays, in complete agreement 
with the experimental observation of \textcite{Wilson:2021}, {\bf iff} the orbital angular 
momentum is not recorded. 
\textcite{Randrup:2021} make a similar claim, which follows 
directly from the assumed form of the FF rotational energy, see Eq.~\eqref{eq:rot}. 
The fact that $\overline{P}_2(S^{L},S^{H})
\approx  \overline{P}_1(S^{L})\overline{P}_1(S^{H})$ (and the corresponding relations 
in the case of other triple distributions)  appears to follow simply from general arguments, 
merely from the conservation of total angular momentum 
$|\langle ({\bm S}^{L} + {\bm S}^{H} +{\bm \Lambda})^2\rangle | = 0$ in case of spontaneous fission, 
with the expected (small) corrections in the case of fission induced by slow neutrons.

The actual spin distribution $P_{\text{true}\,3}(S^{L},S^{H},\Lambda)$ obtained in TDDFT
simulations has a rather complicated structure~\cite{Bulgac:2021b}. 
In Ref.~\cite{Bulgac:2021b} it was shown that 
 \begin{align}
\sum_{ S^{L,H},\Lambda } & |  {\cal N}\hat{P}_1(\Lambda)\hat{P}_1(S^{L}) 
\hat{P}_1(S^{H})\triangle  - P_{\text{true} \,3}(S^{L},S^{H},\Lambda)| \nonumber \\
&=0.35,  \label{eq:3}
\end{align}
where
\begin{align} 
\hat{P}_1(S^{F} )&= \sum_{\Lambda, f\neq  F} {P}_{\text{true} \,3}(S^{F},S^{f},\Lambda),\\
\hat{P}_1(\Lambda) &= \sum_{S^L,S^H}{P}_{\text{true} \,3}(S^{L},S^{H},\Lambda). 
\end{align}
The distributions ${P}_1(S^F)$ used to construct 
$\overline{P}_3(S^{L},S^{H},\Lambda)$ and $\tilde{P}_3(S^L,S^H,\Lambda)$
are slightly different from the distributions $\overline{P}_1(S^F)$
and $\tilde{P}_1(S^F)$ respectively, see Fig.~\ref{fig:TD1}.
In the case of $\overline{P}_3(S^L,S^H,\Lambda)$ 
\begin{align}
\sum_{ S^{L,H},\Lambda } & |  {\cal N}P_1(\Lambda)P_1(S^{L}) 
P_1(S^{H})\triangle  -\overline{P}_{3}(S^{L},S^{H},\Lambda)| \nonumber \\
&\equiv 0   
\end{align} 
by construction.
Using Thomas Dossing's recent suggestion for $\tilde{P}_3(S^{L},S^{H},\Lambda)$ and the corresponding 
$\tilde{P}_1(S^{L,H})$ and $P(\Lambda)$ one obtains 
 \begin{align}
\sum_{ S^{L,H},\Lambda } & |  {\cal N}{P}(\Lambda)\tilde{P}_1(S^{L}) 
\tilde{P}_1(S^{H})\triangle  - \tilde{P}_{3}(S^{L},S^{H},\Lambda)| \nonumber \\
&=0.97.\label{eq:TDF}
\end{align}
In the case of $\overline{P}_3(S^L,S^H,\Lambda)$ I obtain
\begin{align}
\sum_{ S^{L,H},\Lambda } & |  {\cal N}\overline{P}_1(\Lambda)\overline{P}_1(S^{L}) 
\overline{P}_1(S^{H})\triangle  - \overline{P}_{3}(S^{L},S^{H},\Lambda)| \nonumber \\
&=0.28,\label{eq:stF}
\end{align} 
thus closer to the case of the microscopic distribution. 
In this case $\overline{P}_1(S^{L,H})$ and $\overline{P_1}(\Lambda)$ are obtained from 
$\overline{P}_3(S^L,S^H,\Lambda)$ by summing over the other two angular momenta.
The main difference between the distribution suggested by Dossing and the one suggested 
here and the microscopic distribution as well, 
is the weight with which each configuration $(S^L,S^H,\Lambda)$ is accounted for when constructing the
corresponding correlated triple distributions. In the case of Dossing's distribution this weight is
$\propto \left [C^{\Lambda,0}_{S^L,0,S^H,0}\right ]^2/(2\Lambda+1)$ vs 1 in the other cases. Moreover, 
Dossing's distribution allows only even values of $S^L+S^H+\Lambda$~\cite{Varshalovich:1989}, which 
is a rather strong restriction, considering that the FFs emerge highly excited.

An information theory characterization of the distributions discussed in this work is 
through the Shannon measure of the uncertainty, defined for an arbitrary probability distribution as 
$H = - \sum_X [P(X)\log_2 P(X)]$, where $P(X)$ is any probability distribution for a random variable $X$. 
The model distributions $\overline{P}_3(S^L,S^H,\Lambda)$ and 
$ {\cal N}\overline{P}_1(\Lambda)\overline{P}_1(S^{L})  \overline{P}_1(S^{H})\triangle$ 
introduced here have the corresponding uncertainties 12.47 and 12.43 respectively, thus providing a 
comparable amount of uncertainty or information, see Eq.~\eqref{eq:stF} and Figs.~~\ref{fig:P_phi} and~\ref{fig:TD2}.
At the same time, the amount of uncertainty corresponding to the distributions obtained following Dossing's suggestion, 
$\tilde{P}_3(S^L,S^H,\Lambda)$ and 
$ {\cal N}\tilde{P}_1(\Lambda)\tilde{P}_1(S^{L})  \tilde{P}_1(S^{H})\triangle$, 
are 10.71 and 11.45 respectively, which shows that the former has more information, 
in agreement with the Fig.~\ref{fig:TD2}, while the latter has more uncertainty.
 At the same time, as expected,
both probability distributions obtained following Dossing's suggestion have more information, and thus less uncertainty. 
Using the microscopic distribution~\cite{Bulgac:2021b}  $P_{\text{true} \,3}(S^{L},S^{H},\Lambda)$ and  
$ {\cal N}\hat{P}_1(\Lambda)\hat{P}_1(S^{L}) \hat{P}_1(S^{H})\triangle $ one obtains 12.94 and 12.54 
in the case of the SeaLL1 nuclear energy density functional and 12.57 and 11.99 in the case of SkM* respectively.

The probability distribution $p(\phi^{LH})$ however appear to be less sensitive to finer details of the full 
distributions $P_{\text{true} \,3}(S^{L},S^{H},\Lambda)$ from Ref.\cite{Bulgac:2021b},  
$\overline{P}_3(S^{L},S^{H},\Lambda)$ and $\tilde{P}_{3}(S^{L},S^{H},\Lambda)$, before and after the prompt 
neutron and statistical $\gamma$-rays emission, and
it remains a challenge, particularly experimentally, 
to single out the potential observables, which can reveal them.

\section{Conclusions} 

I have presented rather general arguments, not involving
any specific properties of the nuclear interactions, that in a process where a
system with a initial spin $S_0^\pi=0^+$ decays into two fragments
their intrinsic spins form an angle very close to $2\pi/3$ with a
significant dispersion, a feature which one should typically expect in
nuclear spontaneous fission in particular.  The present conclusions agree
with the parameter and assumption free independent microscopic 
calculations performed in Ref.~\cite{Bulgac:2021b}. 
Then, when a nucleus with a very small 
initial spin ${\bm S}_0\approx {\bm 0}$ fissions, the 
distribution of the intrinsic FF spins is determined mainly by statistical 
factors, namely by the rather large number of allowed final values 
of FF intrinsic spins. The range of the allowed FF intrinsic spins,
in particular their distribution,  is controlled 
by their intrinsic deformations~\cite{Randrup:2021,Bulgac:2021,
Bulgac:2021b,Stetcu:2021,Marevic:2021,Vogt:2020}. 
In this respect the FF intrinsic spin distributions 
and their correlations are controlled mostly by the large final 
phase space allowed, as in case of many other types of decays, 
where the allowed phase space, in particular because the fraction of the
phase space volume corresponding to angles  $\phi^{LH}> \pi/2$ 
is approximately 2/3 of the total allowed phase space volume.

Moreover, if one considers the 
reduced distributions of the FF intrinsic spins alone, when the relative orbital
angular momentum $\Lambda$ is not recorded 
\begin{align} 
&\sum _{S^L,S^H} |{P}_2(S^{L},S^{H}) -  {P}_1(S^{L}){P}_1(S^{H})\nonumber 
\\&= 0.08, 0.05\, \text{and}\, 0.02
\end{align}
for the Dossing's distribution $\tilde{P}_3(S^L,S^H,\Lambda)$~\cite{workshop:2022}, the distribution
$\overline{P}_3(S^L,S^H,\Lambda)$ discussed here, and for the microscopic distribution 
$P_{\text{true} \,3}(S^{L},S^{H},\Lambda)$ evaluated in Ref.~\cite{Bulgac:2021b} respectively. 
The distribution $P_2(S^L,S^H)$ appears essentially uncorrelated, even before any 
prompt neutrons or statistical $\gamma$'s are emitted,
thus in full agreement with the observation of \textcite{Wilson:2021}. In this respect 
the system of three angular momenta is similar to the Borromean rings, which fall apart 
if one ring is cut off. The arguments presented here may 
apply also to heavy-ion collisions and atomic and molecular systems as well.    
\vspace{0.3cm}

{\bf Acknowledgements} \\ 

I thank Lee Sobotka for the willingness to have extensive
discussions on these topics and for kindly agreeing to read a few
initial drafts, which helped me sharpen my arguments, G. Scamps for 
discussions and reading of the manuscript, and J. Randrup for a very 
critical reading of the manuscript.  The funding
from the US DOE, Office of Science, Grant No. DE-FG02-97ER41014 and
also the support provided in part by NNSA cooperative Agreement
DE-NA0003841 is greatly appreciated. 
This research used resources of the Oak Ridge
Leadership Computing Facility, which is a U.S. DOE Office of Science
User Facility supported under Contract No. DE-AC05-00OR22725.
 
\section*{Appendix}

It may be useful to consider the distribution $P(S^L,S^H,\phi^{LH})$
instead of the distribution $P_3(S^L,S^H,\Lambda)$. One can show that these two 
distributions are linked in the classical limit by the relation
\begin{align}
&P(S^L,S^H,\phi^{LH})= P_3(S^L,S^H,\Lambda(\phi^{LH})) \frac{ S^LS^H\sin(\phi^{LH}) }{ \Lambda(\phi^{LH})}\nonumber\\
&\Lambda(\phi^{LH}) = \sqrt{(S^L)^2+(S^H)^2+2S^LS^H\cos(\phi^{LH})},\nonumber
\end{align}
and as a result 
\begin{align}
P(\phi^{LH})=\int  dS^L d S^H P(S^L,S^H,\phi^{LH})\propto \sin(\phi^{LH})\nonumber
\end{align}
vanishes exactly at $\phi^{LH} = 0, \pi$, unless the integral 
\begin{align}
\int dS^LdS^h P_3(S^L,S^H,\Lambda(\phi^{LH})) \frac{S^LS^H}{\Lambda(\phi^{LH})}\nonumber 
\end{align}
diverges at least as $1/\sin(\phi^{LH})$. $\phi^{LH}=0$ occurs 
on the upper  face of the pyramid in Fig.~\ref{fig:P3}, when $\Lambda =S^L+S^H$, 
and the two FF intrinsic spins are exactly parallel to each other. $\phi^{LH}=\pi$ occurs
on the lateral faces of the pyramid, when $\Lambda = |S^L-S^H|$ and the two FF intrinsic spins
are anti-parallel. 


\providecommand{\selectlanguage}[1]{}
\renewcommand{\selectlanguage}[1]{}

\bibliography{latest_fission}

\end{document}